\documentclass[showpacs,amsmath,amssymb,twocolumn,nofootinbib]{revtex4}
\begin{document}
\title{Transformations between Jordan and Einstein frames:\\ Bounces,
    antigravity, and crossing singularities}
\author{Alexander~Yu.~Kamenshchik}
\email{kamenshchik@bo.infn.it}
\affiliation{Dipartimento di Fisica e Astronomia, Universit\`a di Bologna\\ and INFN,  Via Irnerio 46, 40126 Bologna,
Italy,\\
L.D. Landau Institute for Theoretical Physics of the Russian
Academy of Sciences,\\
 Kosygin Street 2, 119334 Moscow, Russia}
\author{Ekaterina~O.~Pozdeeva}
\email{pozdeeva@www-hep.sinp.msu.ru}
\affiliation{Skobeltsyn Institute of Nuclear Physics, Lomonosov Moscow State University,\\
 Leninskie Gory 1, 119991, Moscow, Russia}
 \author{Alessandro~Tronconi}
\email{tronconi@bo.infn.it}
\affiliation{Dipartimento di Fisica e Astronomia, Universit\`a di Bologna\\ and INFN, Via Irnerio 46, 40126 Bologna,
Italy}
\author{Giovanni~Venturi}
\email{giovanni.venturi@bo.infn.it}
\affiliation{Dipartimento di Fisica e Astronomia, Universit\`a di Bologna\\ and INFN,  Via Irnerio 46, 40126 Bologna,
Italy}
\author{Sergey~Yu.~Vernov}
\email{svernov@theory.sinp.msu.ru}
\affiliation{Skobeltsyn Institute of Nuclear Physics, Lomonosov Moscow State University,\\
 Leninskie Gory 1, 119991, Moscow, Russia}
\begin{abstract}
We study the relation between the Jordan-Einstein frame transition and the possible description of the crossing of singularities in flat Friedmann Universes, using the fact that the regular evolution in one frame can correspond to crossing singularities in the other frame. We show that some interesting effects arise in simple models such as one  with a massless scalar field or another wherein the potential is constant   in the Einstein frame. The dynamics in these models and in their conformally coupled counterparts are described in detail, and a method for the continuation of such cosmological evolutions beyond the singularity  is developed. We compare
our approach with some other, recently  developed, approaches to the problem of the crossing of singularities.
\end{abstract}
\maketitle

\section{INTRODUCTION}

Models with scalar fields have a special role in modern cosmology. On one hand, they possess a rather rich dynamics and their mathematical properties are very interesting.
On the other hand, the inflationary cosmology \cite{inflation}, describing a very early Universe, uses a scalar field, called inflaton as a matter source of cosmological evolution. Further, the recent discovery of cosmic acceleration  \cite{cosmic} has stimulated the construction of models of dark energy \cite{dark} responsible for this phenomenon.
The dark energy models also use many different kinds of scalar fields as a source of dark energy.

The simplest class of cosmological models, involving scalar fields, are those wherein the scalar field is minimally coupled to gravity. At the same time, more complicated models also became very popular, for example, the models with Born-Infeld type fields,
often called tachyons \cite{BI,tach}, the k-essence fields \cite{k}, etc. Perhaps, the most interesting class of models are the non-minimally coupled scalar fields, whose study descends from the paper by Jordan \cite{Jordan}. In these models the Lagrangian contains a term with the scalar curvature multiplied by a function of the scalar field (usually it is a quadratic polynomial).
The appearance  of such a term is quite natural in the context of the so called induced gravity approach \cite{Sakharov,induced} because quantum corrections to the effective action include such terms \cite{ChernikovTagirov,Callan:1970ze}. Models with non-minimally coupled scalar fields were used in inflationary cosmology \cite{nonmin-inf} and in quantum cosmology \cite{nonmin-quant} and have become especially popular in connection with the so called Higgs inflation \cite{Higgs}.

It is well known that on combining conformal transformation of the metric with the reparametrization of the scalar field, one can rewrite the action of a model with a
non-minimally coupled scalar field in a form where it becomes minimally coupled.
Such a procedure is called the transformation from the Jordan frame to the Einstein frame.
For the first time, this transformation was used in the seminal paper by Wagoner~\cite{Wagoner}.

Many papers discuss this topic, which sometimes is described as a study of the equivalence between the frames \cite{debate,Sasaki}. In a way, one can say that mathematically the procedure of the transition between the frames is well defined and can be used in  different contexts. For example, in a recent paper \cite{Fre} a wide class of the exactly solvable flat Friedmann models with a minimally coupled scalar field was studied. In our preceding paper \cite{KPTVV2013} we have shown that using these solutions one can construct the corresponding solutions in models with non-minimally coupled scalar fields of two types:
the conformally coupled scalar fields in the presence of the Einstein-Hilbert term and the pure induced gravity models without the Einstein-Hilbert term.  In these two cases  the transformation of the scalar field is invertible, and all the formulae are explicit.

We wish to emphasize that the physical cosmological evolutions are those seen by an observer using the cosmic (synchronous) time, which is different in different frames. Thus, evolutions in the Einstein and Jordan frames, connected by a conformal transformation and by the reparametrization of the scalar field, can be qualitatively different. We have constructed an explicit example of such a difference \cite{KPTVV2013}. More precisely, we have considered a de Sitter expansion in induced gravity with a scalar field squared self-interaction potential~\cite{we-ind-ex} and shown that its counterpart is the well-known particular power-law solution~\cite{exp-part} in a minimally coupled model with an exponential potential. We think that while the general solutions for some non-minimally coupled models can be obtained from their minimally coupled counterparts, the study of their behavior is of interest because it can be physically different. The fact that the expansion in one frame can correspond to the contraction in another frame has also been noticed in the literature~\cite{Gasp,Pol1}.

Generally, there are more known exact solutions for cosmological models with minimally coupled scalar fields than with non-minimally coupled ones, and one can think that such solutions are usually simpler. However, it is not always so. In a recent paper~\cite{Boisseau:2015hqa} a flat Friedmann model was considered, where a scalar field was conformally coupled to gravity and the potential included a positive cosmological constant and a negative quartic self-interaction. Such a model has two interesting features: For a large class of initial conditions, the cosmological evolution possesses a bounce and is described in terms of elementary functions. If the initial conditions are chosen for this class, the cosmological evolution (the dependence of the Hubble parameter on cosmic time) does not depend on the evolution of the scalar field, which, in turn, can be described in terms of elliptic functions. In fact, in paper \cite{Boisseau:2015hqa} a narrow class of initial conditions was
  considered, namely, initial conditions which guarantee an evolution of the scalar field such that the coefficient in front of the Ricci scalar in the action, and hence, an effective Newton constant does not change its sign and is positive.

In the following paper \cite{Pol1}, it was shown that using the transition to the Einstein frame
one can convert the model of \cite{Boisseau:2015hqa} into a model with a minimally coupled scalar field with a particular potential, including a fourth degree of the hyperbolical functions. It is interesting that this model, studied in great detail in a series of papers
\cite{Bars1}--\cite{Bars7}, is exactly solvable, and both the cosmological evolution and the evolution of the scalar field are given in terms of elliptic functions. It is important to notice that in papers \cite{Bars1}--\cite{Bars7} the space of evolutions is complete and includes those which undergo a crossing of Big Bang -- Big Crunch singularities and a transition between gravity and antigravity regimes.

The crossing of the so called soft cosmological singularities \cite{soft} was considered in the
literature \cite{cross,cross-we} and does not appear too strange, while the crossing of Big Bang and Big Crunch singularities considered in papers \cite{Bars1}--\cite{Bars7} has provoked some interesting discussions \cite{polem,Bars7}. Let us note that the model in the Jordan frame has a bounce solution~\cite{Boisseau:2015hqa}, whereas the corresponding model in the Einstein frame can describe inflation~\cite{Bars1} (see also~\cite{Roest}).

In our preceding paper \cite{KPTVV2015} we continued the study of the relations between the integrability of the cosmological models in Jordan and Einstein frames, while giving a special emphasis to models where bounces are present.
Here,  continuing the line of research of papers \cite{KPTVV2013,KPTVV2015} and inspired by papers \cite{Boisseau:2015hqa,Pol1} and \cite{Bars1}--\cite{Bars7}, we study the relation between the  Jordan-Einstein frame transition and the possible description of the crossing of singularities. We  use the fact that the regular evolution in one frame can correspond to crossing the singularities in another frame. We show that some interesting effects arise already in models simpler than the one considered in \cite{Boisseau:2015hqa,Pol1} and \cite{Bars1}--\cite{Bars7}.
We wish to emphasize that in the present paper we have considered the case of a flat Friedmann Universe, where the cosmological singularity is isotropic. In papers \cite{Bars3,Bars5,Bars7}, the authors have considered a more general and more complicated case of  anisotropic singularities. We plan to examine the problem of  anisotropic singularities in a future publication.

The structure of the paper is as follows: in Sect. II we recall formulae that connect models in the Einstein and Jordan frames; in the third section we consider the simplest models with a minimally coupled massless scalar field and a vanishing potential, and that with a constant potential.  The fourth section is devoted to the model with the hyperbolic potential \cite{Bars1}--\cite{Bars7} and its conformally coupled counterpart \cite{Boisseau:2015hqa,Pol1}. In the last Section we discuss some questions associated with the crossing of singularities in General Relativity and cosmology and compare our approach with the approach, developed in papers \cite{Bars1}--\cite{Bars7},\cite{Bars8,Bars9}, and that presented in \cite{Wetterich1}--\cite{Wetterich5}.

\section{RELATIONS BETWEEN  MODELS WITH MINIMALLY COUPLED AND NON-MINIMALLY COUPLED SCALAR FIELDS}
\label{Sec2}
Let us consider a cosmological model with the following action
\begin{equation}
S =\int d^4x\sqrt{-g}\left[U(\sigma)R - \frac12g^{\mu\nu}\sigma_{,\mu}\sigma_{,\nu}+V(\sigma)\right],
\label{action}
\end{equation}
where $U(\sigma)$ and $V(\sigma)$ are differentiable  functions of the scalar field $\sigma$.

In a spatially flat Friedman--Robertson--Walker space-time with the interval
\begin{equation}
ds^2 = N^2(\tau)d\tau^2 -a^2(\tau)\vec{dl}^2,
\label{Fried}
\end{equation}
where $a(\tau)$ is the scale factor and $N(\tau)$ is the lapse function,
 one obtains the following equations~\cite{KPTVV2013,KKhT,we-ind-ex}:
\begin{eqnarray}
6Uh^2+6U'\dot{\sigma}h=\frac12\dot{\sigma}^2+N^2V\,,
\label{Fried1}
\end{eqnarray}
\begin{eqnarray}
 &&4U\dot{h}+6Uh^2+4U'\dot{\sigma}h-4Uh\frac{\dot{N}}{N}+2U''\dot{\sigma}^2
\nonumber \\
&&+2U'\ddot{\sigma}-2U'\dot{\sigma}\frac{\dot{N}}{N}= -\frac12\dot{\sigma}^2+N^2V\,,
\label{Fried2}
\end{eqnarray}
\begin{equation}
\ddot{\sigma}+\left(3h-\frac{\dot{N}}{N}\right)\dot{\sigma} -6U'\left[\dot{h}+2h^2-h\frac{\dot{N}}{N}\right]+N^2V' = 0\,,
\label{KG}
\end{equation}
where $h\equiv\dot a/a$, a ``dot'' means a derivative with respect to time, and a ``prime'' means a derivative with respect to $\sigma$.
If we fix the lapse function $N = 1$, then $\tau$ is the cosmic time $t$ and $h$ is the Hubble parameter $H$.

Let us make a conformal transformation of the metric
\begin{equation}
g_{\mu\nu} = \frac{U_1}{U}\tilde{g}_{\mu\nu},
\label{conf}
\end{equation}
where $U_1$ is a constant.
We also introduce a new scalar field $\phi$ such that
\begin{equation}
\frac{d\phi}{d\sigma} = \frac{\sqrt{U_1(U+3U'^2)}}{U}
\quad\Rightarrow\quad
\phi =\! \int\! \frac{\sqrt{U_1(U+3U'^2)}}{U} d\sigma.
\label{scal1}
\end{equation}

The action (\ref{action}) then becomes the action for a minimally coupled scalar field:
\begin{equation}
S =\int d^4x\sqrt{-\tilde{g}}\left[U_1R(\tilde{g}) - \frac12\tilde{g}^{\mu\nu}\phi_{,\mu}\phi_{,\nu}+W(\phi)\right],
\label{action1}
\end{equation}
where
\begin{equation}
W(\phi) = \frac{U_1^2 V(\sigma(\phi))}{U^2(\sigma(\phi))}.
\label{poten}
\end{equation}
Let us emphasise that the formulae (\ref{conf})--(\ref{poten}) are valid for any metric.
In the Einstein frame the spatially flat Friedman--Robertson--Walker metric (\ref{Fried}) becomes
\begin{equation}
\label{EinFreimMetric}
ds^2 =\tilde{N}^2(\tau) d\tau^2 - \tilde{a}^2(\tau) \vec{dl}^2,
\end{equation}
where the new lapse function and the new
scale factor  are defined as
\begin{equation}
\tilde{N} = \sqrt{\frac{U}{U_1}}N,\qquad \tilde{a} =  \sqrt{\frac{U}{U_1}}a.
\label{Na}
\end{equation}

The Friedmann equations and the Klein-Gordon equation in the Einstein frame are as follows:
\begin{equation}
6U_1\tilde{h}^2=\frac12\dot{\phi}^2+\tilde{N}^2W,
\label{Fried10}
\end{equation}
\begin{equation}
4U_1\dot{\tilde{h}}+6U_1\tilde{h}^2-4U_1\tilde{h}\frac{\dot{\tilde{N}}}{\tilde{N}}
={} -\frac12\dot{\phi}^2+\tilde{N}^2W,
\label{Fried20}
\end{equation}
\begin{equation}
\ddot{\phi}+\left(3\tilde{h}-\frac{\dot{\tilde{N}}}{\tilde{N}}\right)\dot{\phi}
+\tilde{N}^2W_{,\phi} = 0,
\label{KG0}
\end{equation}
where $\tilde{h} \equiv \dot{\tilde{a}}/{\tilde{a}}$.

There is a relation between the cosmic time in the Einstein frame~$\tilde{t}$ and the cosmic time in the
Jordan frame~$t$:
\begin{equation}
\tilde{t} = \int \tilde{N}(t) dt = \int \sqrt{\frac{U}{U_1}} dt.
\label{trans1}
\end{equation}

Using the cosmic time in both frames, we get the correspondence between the Hubble parameter $H$ in the Jordan frame and the Hubble parameter $\tilde{H}$ in the Einstein frame.
Using  (\ref{Na}), it is easy to show that
\begin{equation}
H=\sqrt{\frac{U}{U_1}}\left(\tilde{H}-\frac12\frac{d\ln U}{d\tilde{t}}\right).
\label{two-h}
\end{equation}

In what follows we choose
\begin{equation}
U_{c}(\sigma) = U_0 - \frac{\sigma^2}{12},
\label{conf-coupl}
\end{equation}
i.e., we consider the case when the coupling is conformal and a nonzero Einstein--Hilbert term $U_0$ is also present.
Taking into account that
\begin{equation}\label{gsh}
U_c+3{U'_c}^2=U_0,
\end{equation}
 we get from~(\ref{scal1})
\begin{equation}
\label{phic}
\phi=\! \int\! \frac{\sqrt{U_1U_0}}{U_c} d\sigma
\end{equation}
and assume that $U_0>0$.
Hence, $\phi$ is real for $U_1>0$ and is imaginary for $U_1<0$. In the models which we shall consider, the imaginary scalar field can be treated as a real scalar field with a phantom kinetic term.
At $U_1>0$ and $U_c>0$, we get
\begin{equation}
\phi = \sqrt{3U_1}\ln \left[\frac{\sqrt{12U_0}+\sigma}{\sqrt{12U_0}-\sigma}\right]
\label{connection_c0}
\end{equation}
and vice versa
\begin{equation}
\sigma = \sqrt{12U_0}\tanh\left[ \frac{\phi}{\sqrt{12U_1}}\right].
\label{connection_c}
\end{equation}

When, $-\infty < \phi < \infty$, from
(\ref{connection_c}) follows that $U_c>0$. When $|\phi| \rightarrow \infty$, we usually have a singularity in the Einstein frame and the value of the field $\sigma$ in the Jordan frame
tends to its limiting value $|\sigma| = \sqrt{12U_0}$, and hence, $U_c\rightarrow0$.

However, from the point of view of the proper dynamics of the field $\sigma$ in the Jordan frame, there is nothing that prevents it from crossing the value $\sigma = \pm \sqrt{12U_0}$. Indeed, we can represent the corresponding system of equations of motion
as a dynamical system  (see, for example,~\cite{ABGV,Pozdeeva2014}),  obtaining
\begin{equation}
\begin{split}
\dot\sigma &=\zeta,\\
\dot\zeta&={}-3H\zeta-\frac{\left(12U_0-\sigma^2\right)V'+4\sigma V}{12U_0},\\
\dot H&={}-\frac{1}{12U_0}\left(2\sigma^2H^2+\left[4H\zeta-V'\,\right]\sigma+2\zeta^2\right).
\end{split}
\label{FOSEQU}
\end{equation}
It is easy to see that the point where $U_c=0$ is not a singular point of this system.
As was already mentioned in the Introduction, we use that fact to describe
the crossing of the singularities in the Einstein frame, corresponding to the change of the sign of $U_c$ in the Jordan frame. Conversely, we see that the singular behaviour of the Universe in the Jordan frame corresponds  its regular behaviour in the Einstein frame.
A detailed analysis of these facts by using relatively simple models represents the main content of the rest of the paper.

\section{Models with a constant potential}

\subsection{Basic equations}
Let us consider a very simple and well-known cosmological model with a minimally coupled scalar field and a constant potential $W=\Lambda$. It is convenient to work with a cosmic time $\tau = \tilde{t}$. Thus, we fix the lapse function as $\tilde{N} =1$.
Equations (\ref{Fried10})--(\ref{KG0}) take the following form:
\begin{equation}
6U_1\tilde{H}^2=\frac12\dot{\phi}^2+\Lambda,
\label{Fried-mass}
\end{equation}
\begin{equation}
4U_1\dot{\tilde{H}}+6U_1\tilde{H}^2={}-\frac12\dot{\phi}^2+\Lambda,
\label{Fried-mass1}
\end{equation}
\begin{equation}
\ddot{\phi}+3\tilde{H}\dot{\phi}=0.
\label{KG-mass}
\end{equation}

Summing Eqs.~(\ref{Fried-mass}) and (\ref{Fried-mass1}), we obtain
\begin{equation}
\dot{\tilde{H}}+3\tilde{H}^2-\frac{\Lambda}{2U_1}=0.
\label{Fried300m}
\end{equation}
This equation does not include the scalar field and is integrable. Therefore the system of equations (\ref{Fried-mass})--(\ref{KG-mass}) is integrable. Let us note that these equations for the case of 'antigravity' when $U_1<0$ are equivalent to the equations for $U_1>0$ with a phantom scalar fields and $W=-\Lambda$. The explicit form of the solutions depends on the sign of $\Lambda$ (see, for example~\cite{AJV2}).

The conformal transformation combined with the reparametrization of the scalar field, described in Section~\ref{Sec2}, gives the model described by  the action  (\ref{action}) with the following potential
\begin{equation}
V(\sigma)=\frac{\Lambda}{U_1^2}U_{c}^2.
\label{nonmin-neg}
\end{equation}
The Friedmann and Klein-Gordon equations in the Jordan frame [Eqs.~(\ref{Fried1})--(\ref{KG})] with $N=1$  have the following form:
\begin{equation}
6U_{c}H^2+6U_{c}'H\dot{\sigma}=\frac12\dot{\sigma}^2+\frac{\Lambda}{U_1^2} U_{c}^2,
\label{Fried1000}
\end{equation}
\begin{equation}
4U_{c}\dot{H}+6U_{c}H^2+4U_{c}'H\dot{\sigma}+2U_{c}''\dot{\sigma}^2+2U_{c}'\ddot{\sigma}=\frac{\Lambda}{U_1^2}U_{c}^2-\frac12\dot{\sigma}^2,
\label{Fried2000}
\end{equation}
\begin{equation}
\ddot{\sigma}+3H\dot{\sigma}-6U_{c}'\left(\dot{H}+2H^2\right)+2\frac{\Lambda U_{c}'U_{c}}{U_1^2}=0.
\label{KG000}
\end{equation}

On using the explicit form of the potential $V$ and (\ref{Fried1000}), we transform the system~(\ref{FOSEQU}) to a simpler form
\begin{equation}
\ddot{\sigma}={}-3H\dot{\sigma}\,,
\label{KG0001}
\end{equation}
\begin{equation}
\dot{H}+2H^2-\frac{\Lambda}{3U_1^2}\left(U_0-\frac{\sigma^2}{12}\right)=0\,.
\label{Fried2001}
\end{equation}

From Eq.~(\ref{KG0001}) we see that  the scalar field $\sigma$ behaves as a massless scalar field, i.e. its time derivative with respect to the corresponding cosmic time is inversely proportional to the cube of the cosmological
factor: $\dot{\sigma} \sim 1/a^3$.

The field $\sigma$ enters into the second Friedmann equation (\ref{Fried2001}), and this makes the explicit  solution of the system of Eqs.~(\ref{KG0001}) and (\ref{Fried2001}) rather complicated. To integrate the system we subtract Eq.~(\ref{Fried1000}) from Eq.~(\ref{Fried2000}) and get
\begin{equation}
\label{Frieq21}
4U_{c}\dot{H}-2HU_{c}'\dot{\sigma}+2U_{c}''\dot{\sigma}^2+2U_{c}'\ddot{\sigma}+\dot{\sigma}^2=0.
\end{equation}

Thus, we get a system of two equations (\ref{KG0001}) and (\ref{Frieq21}) that do not include the parameter $\Lambda$. This system has the following solution by quadratures:
\begin{equation}
\label{sigmagensol}
t-t_0=\int \frac{d\tilde{\sigma}}{4C_1U_0+36C_2U_0\tilde{\sigma}+C_1\tilde{\sigma}^2+C_2\tilde{\sigma}^3},
\end{equation}
where $t_0$, $C_1$, and $C_2$ are arbitrary constants.
After integration we get
\begin{equation}
\label{sigmagensolm}
   t-t_0=\sum_{i=1}^3\frac{\ln(\sigma-\tilde{\sigma}_i)}{3C_2\tilde{\sigma}_i^2+2C_1\tilde{\sigma}_i+36U_0C_2},
\end{equation}
where $\tilde{\sigma}_i$ are roots of the following cubic equation:
\begin{equation*}
4U_0C_1+36U_0C_2\tilde{\sigma}+C_1\tilde{\sigma}^2+C_2\tilde{\sigma}^3=0.
\end{equation*}

For arbitrary values of the integrable constants $C_1$ and $C_2$, it is difficult to get $\sigma(t)$. For this reason, we consider the Hubble parameter as a function of $\sigma$ (see~\cite{KTVV2013}).
Formula (\ref{sigmagensol}) can be rewritten in the following form:
\begin{equation}\label{dsigmasigma}
    \dot\sigma=4C_1U_0+36C_2U_0\sigma+C_1\sigma^2+C_2\sigma^3.
\end{equation}

Using (\ref{dsigmasigma}), we get the corresponding Hubble parameter
\begin{equation}
\label{Hsol}
H(t)={}-\frac{\ddot\sigma}{3\dot\sigma}={}-12C_2U_0-\frac23C_1\sigma-C_2\sigma^2.
\end{equation}

On substituting~(\ref{dsigmasigma}) and~(\ref{Hsol}) into Eq.~(\ref{Fried1000}), we find
the following relation between the integration constants $C_1$ and $C_2$ and the cosmological constant $\Lambda$:
\begin{equation}\label{LambdaC}
    \Lambda= {}-8U_1^2\left(C_1^2-108U_0C_2^2\,\right).
\end{equation}

For some values of constants $C_1$ and $C_2$, particular solutions can be found in an analytic form.
 The general solution that corresponds to $\Lambda=0$ are presented in the next subsection. For a negative $\Lambda$, it is easy to get particular solutions in an analytic form. For example, at $C_2=0$, we get the following solutions:
\begin{equation*}
\begin{split}
H(t)&={}-\frac{4C_1\sqrt{U_0}}{3} \tan\left(2C_1\sqrt{U_0}(t-t_1)\right),\\
\sigma(t)&={}\pm 2\sqrt{U_0}\tan\left(2C_1\sqrt{U_0}(t-t_1)\right),
\end{split}
\end{equation*}
where $t_1$ is an arbitrary constant and the nonzero parameter $C_1$ is connected with $\Lambda$ by (\ref{LambdaC}), thus $\Lambda<0$.
We see that at some finite moments of time $t$ the function $U_c$ changes the sign on these solutions.

The function $U_c(\sigma)=0$ at $\sigma=\pm\sqrt{12U_0}$. In the neighbourhood of these points, we get the following Taylor series of the general solutions:
\begin{equation}
\label{SolTaipl}
\begin{split}
\sigma& =\sigma_0+16U_0\left(C_1+3C_2\sigma_0\right)t+{\cal{O}}(t^2),\\
H &={}-2C_2\sigma_0^2-\frac{2}{3}\sigma_0 C_1\\
&{}-\frac{32}{3}U_0\left[C_1+3C_2\sigma_0\right]^2t+{\cal{O}}(t^2).
\end{split}
\end{equation}
where $\sigma_0=\pm\sqrt{12U_0}$, and we assume that $U_c=0$ at $t=0$.

We can also make a useful general observation from the study of Eq. (\ref{Fried1000}) at the moment $t=0$, when
$U_c=0$ and $\sigma = \sigma_0$. From (\ref{dsigmasigma}) and (\ref{Hsol}), it is easy to see that
\begin{equation}
\dot{\sigma}(0) = {}-2\sigma_0H(0).
\label{Fried-rel}
\end{equation}
From here, it follows that when the field $\sigma$ and its time derivative have the same sign, the Hubble parameter is negative, and when they have the opposite signs, the Hubble parameter is positive. Obviously, when these quantities have the same sign the field $\sigma$ leaves the region where $U_c > 0$ and enters into the region, where $U_c < 0$. We see that in the Einstein frame this situation corresponds to the Big Crunch singularity  and $\tilde{H} \rightarrow -\infty$. To cross this singularity by using the correspondence between the Jordan and Einstein frames, we should make the transition from the gravity regime to the antigravity regime. If instead the field $\sigma$ and its time derivative $\dot{\sigma}$ have the opposite signs, which implies the positivity of the Hubble parameter, the field $\sigma$ enters into the regime $U_c>0$, while its counterpart in the Einstein frame enters into the gravity regime from the antigravity regime on crossing the Big Bang
singularity.

In the rest of this section we shall consider in some detail the cases of zero and  of the negative cosmological constant. The case for positive $\Lambda$ is technically quite analogous to the case of the negative one, but the variety of the cosmological evolutions is less rich. Thus, we do not dwell on this case.

\subsection{Massless scalar field}

At $\Lambda=0$, the general solution of Eq.~(\ref{Fried300m}) is
\begin{equation}
\tilde{H} = \frac{1}{3(\tilde{t}-\tilde{t}_1)}.
\label{massless1}
\end{equation}
We can assume that the integration constant $\tilde{t}_1=0$ without the loss of generality. The corresponding evolution of the scale factor is
\begin{equation}
\tilde{a}(\tilde{t}) = \tilde{a}_0{\tilde{t}}^{1/3}.
\label{cosm-mass}
\end{equation}

Solution (\ref{massless1}) describes an expansion ($0 < \tilde{t} < \infty$)  or a contraction
($-\infty < \tilde{t} < 0$) of the flat Friedmann Universe filled with stiff matter having the equation of state $p = \rho$, where $p$ is the pressure and $\rho$ is the energy density.

The expansion begins from an initial singularity --- Big Bang, while the contraction ends
with the final singularity --- Big Crunch. These two evolutions are totally disconnected and do not depend on the details of the evolution of the scalar field.

Substituting the solution (\ref{massless1}) into Eq.~(\ref{KG-mass}), we find
\begin{equation}
\phi = \phi_1\ln|\tilde{t}| + \phi_0,
\label{cosm-mass1}
\end{equation}
where $\phi_1$ and $\phi_0$ are constants. On substituting the solution (\ref{cosm-mass1}) into Eq.~(\ref{Fried1000}) we find
\begin{equation}
\phi = \pm 2\sqrt{\frac{U_1}{3}} \ln|\tilde{t}| + \phi_0.
\label{cosm-mass3}
\end{equation}
We note that the constants $\phi_0$ and $\tilde{a}_0\neq 0$ are arbitrary.

Let us consider the corresponding model in the Jordan frame.
Equation~(\ref{Fried2001}) with $\Lambda=0$
has the following general solution:
\begin{equation}
H=\frac{1}{2(t-t_1)},
\label{rad1}
\end{equation}
which describes the expansion or the contraction of a flat Friedmann Universe filled with radiation. Again its evolution does not depend on the details of the evolution of the scalar field. We can set an integration constant $t_1=0$ without loss of the generality. The scale factor behaves as follows:
\begin{equation}
a=a_0\sqrt{|t|}
\label{rad-conf}
\end{equation}
and the scalar curvature $R=0$.
On substituting the expression (\ref{rad1}) into Eq.~(\ref{KG0001}) we obtain an equation
\begin{equation}
\ddot{\sigma}+\frac{3}{2t}\dot{\sigma}=0.
\label{KG-conf1}
\end{equation}
Its solution is
\begin{equation}
 \sigma = \frac{\sigma_1}{\sqrt{|t|}}+\sigma_0.
 \label{KG-conf2}
 \end{equation}
On substituting this solution into the Friedmann equation (\ref{Fried1000}) with $\Lambda=0$, we find that
 \begin{equation}
 \sigma=\frac{\sigma_1}{\sqrt{|t|}} \pm \sqrt{12U_0}.
 \label{KG-conf3}
 \end{equation}
We see that the function $\sigma(t)$ has a singular point at $t=0$, hence, it is more correct to write the solution (\ref{KG-conf3}) in the following form:
\begin{equation*}
 \sigma=\left\{
 \begin{split}
 \frac{\tilde{\sigma}_1}{\sqrt{-t}} \pm \sqrt{12U_0},&\quad t<0,\\
 \frac{\sigma_1}{\sqrt{t}} \pm \sqrt{12U_0},&\quad t>0.
 \end{split}
 \right.
\end{equation*}
The only restrictions on the values of the constants are
$\tilde{\sigma}_1\neq 0$, $\sigma_1\neq 0$, and $a_0\neq 0$. We show that it is possible to connect $\tilde{\sigma}_1$ with $\sigma_1$ using
the corresponding solution in the Einstein frame.

To analyze the behavior of the solution we should consider different cases, depending on the sign choice for the constants $\sigma_0$ and $\sigma_1$. The equations are invariant with respect to the change $\sigma(t)$ to $-\sigma(t)$. So, we can assume that $\sigma_0>0$ which means $\sigma_0=\sqrt{12U_0}$  without loss of generality and consider separately positive and negative $\sigma_1$. Solutions with $\sigma_1>0$ correspond to $U_c<0$ for any value of $t$, whereas solutions with $\sigma_1<0$ correspond to a $U_c$ that changes sign at $t = \sigma_1^2/(48U_0)$.

Let us now analyse the relations between the evolutions in the minimally and non-minimally coupled models by using the correspondence formulas from the Section~\ref{Sec2}. Firstly, we  consider the case $\sigma_1>0$ and time running from $0$ to $\infty$.
The fact that  in this case $U_c$ is always negative is not important from the point of view of the evolution of the Hubble parameter and of the scale factor in the Jordan frame.
However, we would like to have both the scale factor and the lapse function real in the Einstein frame also. On looking at the correspondence formulae (\ref{Na}), we see that this requires $U_1 < 0$, i.e., in the Einstein frame, and instead of gravity we would have antigravity. Then, on substituting
$U_1 < 0$ into Eq.~(\ref{scal1}) describing the transition between the scalar fields in the
Jordan and Einstein frames, we see that the field $\phi$ becomes imaginary. This is equivalent
to changing the sign of the kinetic term in the Lagrangian. In this model with zero potential the simultaneous substitution of gravity by antigravity and of the standard scalar field with the phantom scalar field leaves the Friedmann equations invariant.

It is convenient to introduce another notation for the phantom scalar field:
\begin{equation}
\chi=\sqrt{3|U_1|}\ln\frac{\sigma+\sqrt{12U_0}}{\sigma-\sqrt{12U_0}}.
\label{chi}
\end{equation}
On inverting
\begin{equation}
\sigma=\sqrt{12U_0}\coth\frac{\chi}{\sqrt{12|U_1|}}.
\label{chi1}
\end{equation}

On using~(\ref{trans1}), we explicitly get
\begin{equation}
\tilde{t}=\frac{(\sigma_1^2+2\sqrt{12U_0}\sigma_1\sqrt{t})^{3/2}}{18\sigma_1\sqrt{U_0|U_1|}}
\label{trans2}
\end{equation}
and on inverting,
\begin{equation}
t=\left[\frac{(18\sqrt{U_0|U_1|}\sigma_1\tilde{t})^{2/3}-\sigma_1^2}{2\sqrt{12U_0}\sigma_1}\right]^2.
\label{trans3}
\end{equation}

On using formulae (\ref{chi}) and (\ref{trans3}), we find
the dependence of the phantom field on the cosmic time $\tilde{t}$:
\begin{equation}
\chi=2\sqrt{\frac{|U_1|}{3}}\ln\frac{\tilde{t}}{\tilde{t}_0},
\label{chi2}
\end{equation}
where
\begin{equation}
\tilde{t}_0=\frac{\sigma_1^2}{18\sqrt{|U_1|U_0}}.
\label{chi3}
\end{equation}

Let us now consider the cosmological expansion in the Jordan frame.
 At the instant $t = 0$, the value of the cosmological factor
in the Einstein frame can be obtained by using the formulae (\ref{Na})
and (\ref{KG-conf3}). We then have
\begin{equation}
\tilde{a} = \lim_{t\rightarrow 0_+} \sqrt{\frac{U}{U_1}}a(t) = a_0\frac{\sigma_1}
{\sqrt{12|U_1|}},
\label{trans}
\end{equation}
i.e., the scale factor in the Einstein frame is finite at the moment
when in the Jordan frame we have the Big Bang singularity.
One can see that the instant $t=0$, when we encounter the   Big Bang  in the Jordan frame, corresponds to
a time $\tilde{t} = \tilde{t}_0$ in the Einstein frame.
After this time, the evolution of the scale factor $\tilde{a}$ in the Einstein frame is given by
\begin{equation}
\frac{a_0(18\sqrt{U_0|U_1|}\sigma_1)^{1/3}}{\sqrt{12|U_1|}}\tilde{t}^{1/3}.
\label{trans4}
\end{equation}
However, this expression is perfectly valid for $\tilde{t} < \tilde{t}_0$ also.
The same is true also for the evolution of the phantom field~(\ref{chi2}).

On now using the continuity of the evolution in the Einstein frame, we can describe
the crossing of the Big Bang singularity in the Jordan frame.
First of all, using formula (\ref{chi1}), we can find the relation between $t$ and $\tilde{t}$ at $\tilde{t} \leq \tilde{t}_0$:
\begin{equation}
t=-\frac38\sqrt{\frac{|U_1|}{U_0}}\tilde{t}_0\left[\left(\frac{\tilde{t}}{\tilde{t}_0}\right)^{2/3}-1\right]^2.
\label{trans5}
\end{equation}
Naturally, the negative values of the cosmic time $t$ correspond to the contraction of the Universe [see formula~(\ref{rad1})].
On inverting,
\begin{equation}
\tilde{t}=\tilde{t}_0\left(1-\sqrt{\frac{t}{t_0}}\right)^{3/2},
\label{trans6}
\end{equation}
where
\begin{equation}
t_0=-\frac{1}{48}\frac{\sigma_1^2}{U_0}.
\label{trans7}
\end{equation}

On now using the formulae (\ref{trans6}) and (\ref{chi1}), we find that
for $t < 0$, the scalar field $\sigma$ behaves as
\begin{equation}
\sigma(t) = \sqrt{12U_0}-\frac{\sigma_1}{\sqrt{-t}}.
\label{trans8}
\end{equation}
Thus, when $-\infty < t < 0$, we have a contraction of the Universe in the Jordan frame while the scalar field is given by (\ref{trans8}).
In other words, we get that $\tilde{\sigma}_1=-\sigma_1$. This means that $\tilde{\sigma}_1<0$, hence, $U_c=0$ at some moment. It is easy to see that at the instant $t=t_0$ the scalar field $\sigma = -\sqrt{12U_0}$. Hence, the function $U$ becomes positive in the Jordan frame and we make a transition from antigravity to gravity in the Einstein frame.
On the other hand, at the instant when $t=t_0$, the cosmic time in the Einstein frame is equal to zero: $\tilde{t} =0$, and the Universe encounters the Big Bang singularity. However, on using the fact that the evolution in the Jordan frame is regular, we can pass through this singularity to the region where $\tilde{t} < 0$ [see formula (\ref{trans6})].  Thus, on passing through the Big Bang singularity, the constant $U_1$ changes sign and the phantom field transforms into a standard scalar field $\phi$ connected with the field $\sigma$ through  the relations (\ref{connection_c0}) and (\ref{connection_c}).
Finally, when $t<t_0$ and, correspondingly, $\tilde{t} < 0$, we have a contraction of the Universe in both frames.

Let us note that if instead of choosing $\sigma_1 > 0, \sigma_0 = +\sqrt{12U_0}$ we had chosen for
both different signs, the qualitative picture of the evolutions in the Einstein and Jordan frames would not be changed. In any case, we have in both frames the contraction of the Universe until the Big Crunch singularity with the subsequent Big Bang, when the Universe begins its expansion. The point is that the instants of encounter with  the singularity do not coincide in the two frames, and this gives us the opportunity of crossing the singularity in one frame using the smoothness of the evolution in the other frame.
This crossing of the singularity corresponds to the singularity crossing described in the papers by Bars \textit{et al.} for more complicated models by using a slightly different mechanisms for the changes of the parametrization of the gravitational and scalar fields.

\subsection{The model with a negative cosmological term}

In this section we consider a more complicated and rich cosmological model, which
in the Einstein frame appears as a model with a massless scalar field and a negative cosmological term, i.e., a model with the potential
\begin{equation}
W(\phi) =\Lambda={} -\bar{\Lambda},\qquad \bar{\Lambda} > 0.
\label{neg}
\end{equation}

From Eq.~(\ref{Fried300m}), we obtain
\begin{equation}
\dot{\tilde{H}}+3\tilde{H}^2+\frac{\bar{\Lambda}}{2U_1}=0,
\label{Fried300}
\end{equation}
therefore,
\begin{equation}
\tilde{H}(\tilde{t})=\sqrt{\frac{\bar{\Lambda}}{6U_1}}\cot\left[\sqrt{\frac{3\bar{\Lambda}}{2U_1}}(\tilde{t}-\tilde{t}_1)\right].
\label{Fried400}
\end{equation}
Let us consider a solution with $\tilde{t}_1=0$.
This Hubble parameter describes an evolution of the flat Friedmann Universe,
which begins at the Big Bang singularity at the instant $\tilde{t}=0$, arrives at the point of maximal expansion at $\tilde{t}=\frac{\pi}{2}\sqrt{2U_1/\left(3\bar{\Lambda}\right)}$, and then contracts until the encounter with the Big Crunch singularity at $\tilde{t}=\pi \sqrt{2U_1/\left(3\bar{\Lambda}\right)}$. Let us note that this evolution does not depend on the evolution of the scalar field $\phi$.  On now substituting the function (\ref{Fried400}) into the Klein-Gordon equation
(\ref{KG-mass}) we find
\begin{equation}
\dot{\phi}(\tilde{t}) =\frac{\phi_1}{\sin\left[\sqrt{\frac{3\bar{\Lambda}}{2U_1}}\tilde{t}\right]}.
\label{KG001}
\end{equation}
On substituting this expression into the first Friedmann equation (\ref{Fried-mass}) we obtain
\begin{equation}
\phi_1 =\pm\sqrt{2\bar{\Lambda}}
\label{phi-1}
\end{equation}
Finally, on integrating (\ref{KG001}) and taking into account (\ref{phi-1}) we obtain
\begin{equation}
\phi(\tilde{t}) = \pm 2\sqrt{\frac{U_1}{3}}\ln\left[\tan\sqrt{\frac{3\bar{\Lambda}}{8U_1}}\tilde{t}\right]+\phi_0,
\label{KG002}
\end{equation}
where $\phi_0$ is an integration constant.

The conformal transformation combined with the reparametrization of the scalar field, described in Section~\ref{Sec2}, gives the model with the action  (\ref{action}), where the function $U$  corresponds to the conformal coupling (\ref{conf-coupl}) and the potential is given by (\ref{nonmin-neg})
\begin{equation}
V(\sigma)={}-\bar{\Lambda}\frac{U_c(\sigma)^2}{U_1^2}.
\label{nonmin-neg2}
\end{equation}
Let us note that models with non-minimally coupled scalar fields and potentials that are not positive definite are actively studied~\cite{Boisseau:2015hqa,KPTVV2015,ABGV}, in particular, to obtain bounce solutions.

Let us now consider the evolution of the Universe in the Einstein frame from the Big Bang to the Big Crunch (\ref{Fried400}) and the corresponding evolution in the Jordan frame. Choosing the ``plus'' sign in the expression (\ref{KG002}) for definiteness, we get that the scalar field $\phi$ changes from $-\infty$ to $+\infty$ on the interval $0<\tilde{t}<\pi \sqrt{2U_1/\left(3\bar{\Lambda}\right)}$, and hence, due to formula (\ref{connection_c}), the field $\sigma$ runs from $-\sqrt{12U_0}$ to $+\sqrt{12U_0}$.

Let us calculate the Hubble parameter in the Jordan frame at the point $U_c=0$. To do this, we use (\ref{two-h}) and the following formulae. The function $U$, expressed in terms of the scalar field $\phi$, is
\begin{equation}
U=\frac{U_0}{\cosh^2\left(\frac{\phi}{\sqrt{12U_1}}\right)}.
\label{U-phi}
\end{equation}
On using the explicit dependence of the field $\phi$ on the cosmic time $\tilde{t}$, we obtain
\begin{widetext}
\begin{equation}
\sqrt{\frac{U}{U_1}}=2\sqrt{\frac{U_0}{U_1}}\frac{1}{\left(\tan\sqrt{\frac{3\bar{\Lambda}}{8U_1}}\tilde{t}\right)^{1/3}\exp \left(\frac{\phi_0}{\sqrt{12U_1}}\right)+\left(\tan\sqrt{\frac{3\bar{\Lambda}}{8U_1}}\tilde{t}\right)^{-1/3}\exp\left( -\frac{\phi_0}{\sqrt{12U_1}}\right)}.
\label{U-phi1}
\end{equation}
Further
\begin{equation}
-\frac12\frac{d\ln U}{d\tilde{t}}=\frac{d\ln\cosh\frac{\phi}{\sqrt{12U_1}}}{d\tilde{t}}
=\frac{\sqrt{\frac{\bar{\Lambda}}{24U_1}}
\left[\left(\tan\sqrt{\frac{3\bar{\Lambda}}{8U_1}}\tilde{t}\right)^{-2/3}\exp\left(\frac{\phi_0}{\sqrt{12U_1}}\right)
\mathbf{ - } \left(\tan\sqrt{\frac{3\bar{\Lambda}}{8U_1}}\tilde{t}\right)^{-4/3}\exp\left(-\frac{\phi_0}{\sqrt{12U_1}}\right)\right]}
{\cos^2\left(\sqrt{\frac{3\bar{\Lambda}}{8U_1}}\tilde{t}\right){\left[\left(\tan\sqrt{\frac{3\bar{\Lambda}}{8U_1}}\tilde{t}\right)^{1/3}
\exp\left(\frac{\phi_0}{\sqrt{12U_1}}\right)
+\left(\tan\sqrt{\frac{3\bar{\Lambda}}{8U_1}}\tilde{t}\right)^{-1/3}\exp\left(-\frac{\phi_0}{\sqrt{12U_1}}\right)\right]}}.
\label{U-phi2}
\end{equation}
\end{widetext}

On now substituting formulae (\ref{Fried400}), (\ref{U-phi1}) and (\ref{U-phi2}) into
Eq.~(\ref{two-h}), and taking the limit $\tilde{t} \rightarrow 0_+$, we obtain
\begin{equation}
H_0=\lim_{\tilde{t}\rightarrow 0_+}H(\tilde{t}) = 2\sqrt{\frac{U_0\bar{\Lambda}}{6U_1^2}}\exp \frac{3\phi_0}{\sqrt{12U_1}} > 0.
\label{BB}
\end{equation}
Thus, in the Jordan frame, at this point, the evolution is regular, and the Universe is finite and is still expanding.
Using this fact we can describe the crossing of the singularity in the Einstein frame.
However, in contrast with the model described in the preceding subsection, it is difficult to find an explicit relation between the cosmic times in the Einstein and Jordan frames, and we  use an asymptotic analysis of the behaviour of the Universe in the vicinity of singularities.

As we have already said, the singularity of the function $\sigma$ corresponds to singularity of the function $U_c$, i.e., to the singularity in the Jordan frame. The crossing of this singularity can be analysed using the smooth behaviour of solutions in the Einstein frame by the method suggested in the previous subsection.
On the other hand when we have the singularity in the Einstein frame, the scalar field $\phi$ tends to infinity, and in the Jordan frame, $U_c=0$. We describe in some detail the crossing of this singularity by using the regularity of the evolution in the Jordan frame. The crossing of the singularity in the Jordan frame can be described analogously.

We have already written down the relation (\ref{Fried-rel}), which connects the values of the scalar field $\sigma$, its time derivative $\dot{\sigma}$, and the Hubble variable in the Jordan frame at the moment $t=0$, when $U_c$ disappears. If we choose
$\sigma=\sigma_0=-\sqrt{12U_0}$ and $\dot{\sigma}(0)=\sigma_1> 0$, then the Hubble parameter in the Jordan frame is positive
and the field $\sigma$ enters into the region where $U_c> 0$ from the region where $U_c <0$. Correspondingly, the Universe in the Einstein frame crosses the Big Bang singularity making the transition from the antigravity regime to the gravity regime.

Thus, in the vicinity of  $t=0$ the field $\sigma$ behaves as
\begin{equation}
\sigma(t) = {}-\sqrt{12U_0}+\sigma_1 t+{\cal O}(t^2),
\label{BB1}
\end{equation}
 where the value of  $\sigma_1$ can be found from Eqs.~(\ref{Fried-rel}) and (\ref{BB}):
 \begin{equation}
 \sigma_1=8U_0\sqrt{\frac{3}{U_1}}\exp\frac{3\phi_0}{\sqrt{12U_0}}.
 \label{sigma1}
 \end{equation}

 Obviously, for $t < 0$, the expression for $U$ becomes negative, and hence, in the Einstein frame we should substitute gravity by antigravity and the standard scalar field by the phantom. The evolution in the Einstein frame is described by the formula
 \begin{equation}
 \tilde{H}(\tilde{t})=\sqrt{\frac{\bar{\Lambda}}{6|U_1|}}\coth\left(\sqrt{\frac{3\bar{\Lambda}}{2U_1}}\tilde{t}\right),
 \label{contr-BC}
 \end{equation}
 where the time parameter $\tilde{t}$ runs from $-\infty$ to $0$ and one has a contraction which culminates in the encounter with the Big Crunch singularity.
 The phantom field behaves as
 \begin{equation}
 \chi(\tilde{t})=\pm2\sqrt{\frac{|U_1|}{3}}\ln\tanh\left(-\sqrt{\frac{3\bar{\Lambda}}{8|U_1|}}\tilde{t}\right)+\chi_0.
 \label{contr-BC1}
 \end{equation}
 If we choose the sign ``plus'' in the formula above, then the correct
  relation between the field $\sigma$ in the Jordan frame and the phantom $\chi$ is given by
 \begin{equation}
 \sigma=\sqrt{12U_0}\coth\frac{\chi}{\sqrt{12|U_1|}}.
 \label{contr-BC2}
 \end{equation}

We now wish to understand what happens with the Universe in the Jordan frame
 at $-\infty < \tilde{t} < 0$. As we have already mentioned, at the instant $\tilde{t} = 0$, the Universe in the Jordan frame is expanding. Thus, there are three possible evolutions before this instant. One can have an infinite expansion before an initial instant (just like the case of a flat de Sitter Universe, expanding according to the law $a(t) \sim \exp(H_0 t)$). One can have an expansion beginning from the Big Bang type singularity, or one can have in its past a bounce, which was preceded by a contraction. Let us try to analyze which of these scenarios can be realized and at which conditions.

We again use the formula (\ref{two-h}). First of all, let us write down the analogue of the formula (\ref{U-phi2}) for the evolution at $\tilde{t} < 0$:
\begin{widetext}
\begin{eqnarray}
&&-\frac12\frac{d\ln U}{d\tilde{t}}=\frac{d\ln\sinh\frac{\chi}{\sqrt{12|U_1|}}}{d\tilde{t}}\nonumber \\
&&=
\frac{-\sqrt{\frac{\bar{\Lambda}}{24|U_1|}}}{\cosh^2\left(-\sqrt{\frac{3\bar{\Lambda}}{8|U_1|}}\tilde{t}\right)}
\frac{\left[\tanh\left(-\sqrt{\frac{3\bar{\Lambda}}{8|U_1|}}\tilde{t}\right)\right]^{-2/3}
\exp\left(\frac{\chi_0}{\sqrt{12|U_1|}}\right)+\left[\tanh\left(-\sqrt{\frac{3\bar{\Lambda}}{8|U_1|}}
\tilde{t}\right)\right]^{-4/3}\exp\left(-\frac{\chi_0}{\sqrt{12|U_1|}}\right)}
{\left[\tanh\left(-\sqrt{\frac{3\bar{\Lambda}}{8|U_1|}}\tilde{t}\right)\right]^{1/3}
\exp\left(\frac{\chi_0}{\sqrt{12|U_1|}}\right)-\left[\tanh\left(-\sqrt{\frac{3\bar{\Lambda}}{8|U_1|}}
\tilde{t}\right)\right]^{-1/3}\exp\left(-\frac{\chi_0}{\sqrt{12|U_1|}}\right)}.
\label{log-der-chi}
\end{eqnarray}
\end{widetext}
Combining the formulae (\ref{log-der-chi}), (\ref{contr-BC}), and (\ref{two-h}) and taking the limit $\tilde{t} \rightarrow 0_{-}$, we find
\begin{equation}
H_0=\lim_{\tilde{t}\to 0_{-}}H(\tilde{t}) = 2\sqrt{\frac{U_0\bar{\Lambda}}{6U_1^2}}\exp\left(\frac{3\chi_0}{\sqrt{12|U_1|}}\right).
\label{h-0-2}
\end{equation}

Naturally, the right-hand sides of the formulae (\ref{BB}) and (\ref{h-0-2}) should coincide, and hence,
\begin{equation}
\chi_0=\phi_0.
\label{coin-chi-phi}
\end{equation}

We are now in a position to analyse the past evolution of the Universe in the Jordan frame.
First of all, let us note that the opportunity of having a singularity of the Big Bang type for some value of $\tilde{t} < 0$ is conditioned by the vanishing of the function $\sinh\frac{\chi}{\sqrt{12|U_1|}}$.
For
\begin{equation}
 \chi(\tilde{t})=2\sqrt{\frac{|U_1|}{3}}\ln\left[\tanh\left(-\sqrt{\frac{3\bar{\Lambda}}{8|U_1|}}\tilde{t}\right)\right]+\chi_0,
 \label{contr-BC3}
 \end{equation}
it  is possible if and only if $\chi_0$ is positive. What happens when $\chi_0$ is negative?
To answer this question, we can try to find the limit $H(\tilde{t})$ for $\tilde{t}\to -\infty$, again by using the formulae (\ref{log-der-chi}), (\ref{contr-BC}) and (\ref{two-h}):
\begin{equation}
\lim_{\tilde{t} \to -\infty}H(\tilde{t}) = -\sqrt{\frac{\bar{\Lambda} U_0}{6U_1^2\sinh^2\frac{\chi_0}{\sqrt{12|U_1|}}}}.
\label{lim-tilde-inf}
\end{equation}
Thus, we see that at $\tilde{t} \rightarrow -\infty$ the Universe had a contraction with a finite negative value of the Hubble parameter, then, at some moment, it had a bounce, the Hubble parameter changed sign and the expansion began.

If
\begin{equation*}
\phi_0=\chi_0 > 0,
\end{equation*}
then, as we have already mentioned, the Universe had a Big Bang singularity in the Jordan frame, while in the Einstein frame at this instant, it was contracting in a non-singular way. Using this fact, we describe the transition from the contraction to the expansion or the passage through
the Big Crunch--Big Bang singularity by using the regularity of the evolution in the Einstein frame, just as was done in the preceding section for a simpler model with a vanishing potential.

\section{Models with the hyperbolic potentials and their Jordan frame counterparts}

As we have already mentioned in the Introduction, in paper \cite{Boisseau:2015hqa} a model was considered with a conformal coupling, which in our notation is given by formula
(\ref{conf-coupl}), while the potential can be written as
\begin{equation}
V(\sigma)=V_0+V_1\chi^4.
\label{Pol-Star}
\end{equation}
In this case the Friedmann equation governing the cosmological evolution is astonishingly simple
\begin{equation}
6U_0H^2 = V_0+\frac{A}{a^4},
\label{Pol-Star1}
\end{equation}
where the constant $A$ is connected with the scalar field $\sigma$ by
\begin{equation}
A=\frac12\left(\frac{d}{d\eta}\left[\frac{\sigma}{a}\right]\right)^2+V_1\sigma^4.
\label{Pol-Star2}
\end{equation}
and $\eta$ is the conformal time parameter. If the coupling constant $V_1$ of the quartic self-interaction of the field $\sigma$ is negative, then one can choose the initial conditions
for this field in such a way that the integration constant $A$ is also negative and in this case
the cosmological evolution has a  bounce at
\begin{equation}
a_{\rm bounce} = \left(-\frac{A}{V_0}\right)^{1/4}.
\label{Pol-Star3}
\end{equation}
This cosmological evolution is quite regular. At the same time, to the same cosmological evolution, totally characterised by the choice of the parameter $A$, correspond different evolutions for the scalar field $\sigma$.
This family of trajectories can be parameterised, for example, by the
 value of the time derivative of the scalar field at the moment when the Universe undergoes the bounce. During some of these evolutions, the field $\sigma$ takes values which imply a change of sign of $U$. This means that the effective Newton constant changes sign. Such situations were treated in~\cite{Boisseau:2015hqa} as non-physical, and the conditions which allow one to exclude such evolutions were determined.
In paper \cite{Pol1} it was noticed that on making conformal transformations of the metric
and the corresponding reparametrisation
of the scalar field (such as that described in Section II) one arrives to a cosmological model with a minimally coupled scalar field with the potential
\begin{equation}
W(\phi)=\frac{U_1^2}{U_0^2}\left[V_0\cosh^4\frac{\phi}{\sqrt{12U_1}}+144U_0^2V_1\sinh^4\frac{\phi}{\sqrt{12U_1}}\right].
\label{Pol-Star4}
\end{equation}
As we have already mentioned, this model was studied in detail in papers \cite{Bars1}--\cite{Bars7}. In paper
\cite{Pol1} the correspondence between the potentials (\ref{Pol-Star}) and (\ref{Pol-Star4}) was used to render the analysis of the cosmological evolutions in the model with the conformally coupled scalar field more transparent. The evolutions for which the effective Newton constant changes sign were also treated in \cite{Pol1}, where they were still considered as unphysical and unrealistic. The relations between the minimally coupled models of the type (\ref{Pol-Star4}) and the conformally coupled model with the potential (\ref{Pol-Star}) and their place in the general classification of  exactly solvable cosmological models \cite{Fre} was discussed in our preceding paper \cite{KPTVV2015}.

In the series of papers \cite{Bars1}--\cite{Bars7} the dynamics of the model was considered in detail and the question of the crossing of singularities and of the possible evolution in the antigravity regime was analysed by using a special representation of the theory. Namely, instead of one scalar field the authors introduce two scalar fields, say, $\Phi$ and $\Psi$ with their kinetic terms, which have opposite signs and with a conformal coupling to the curvature.
Naturally, the terms describing this conformal couplings of the fields also have the opposite signs.
On choosing the potential term in the form $\Phi^4f(\Phi/\Psi)$, where $f$ is an arbitrary function, the  Weyl invariance of the theory is guaranteed.

One can then represent these two scalar fields in a ``hyperbolic form'':
 \begin{equation}
 \Phi=r\cosh\phi,\qquad \Psi=r\sinh\phi.
 \label{hyp-repr}
 \end{equation}
On using the Weyl invariance of the theory, one can fix the value of $r$. In this case the theory transforms into one with minimally coupled scalar field $\phi$ and with the standard Hilbert--Einstein term. On choosing in a special way the function $f$, one obtains the potential (\ref{Pol-Star4}).
However, one is not obliged to use the parametrization (\ref{hyp-repr}). On using Weyl invariance, one can consider an evolution of the geometric characteristics of the Universe and of the scalar fields $\Phi$ and $\Psi$ such that the Universe passes through the Big Bang -- Big Crunch singularity  and simultaneously gravity is substituted by antigravity (i.e., one has $\Phi^2 - \Psi^2 < 0$) or vice versa.  Formally, this transition to the antigravity regime resembles the crossing of the horizon of the Schwarzschild black hole in Kruskal coordinates~\cite{Kruskal}.

In Section III of the present paper, we have shown how a similar procedure can be pursued by
using the transitions between the Einstein and Jordan frames. Considering simpler models, we have shown that the passing through the singularity in the Einstein frame accompanied by the transition from the gravity to the antigravity regime can, in a nonambiguous way, be described by using the transition to the Jordan frame, where the corresponding scalar field is coupled conformally to gravity.  Moreover, the procedure is also valid when it is applied in the opposite direction: We can describe the passage through the singularity in the Jordan frame by using the transition to the Einstein frame.

Since cosmological evolution in the models with the hyperbolic potentials and in their Jordan frame counterparts has been studied extensively in the papers cited above and we have illustrated our method by the examples of simpler models, here we limit ourselves to one short comment.  One can always find which evolution in the Einstein frame corresponds to a given evolution in the Jordan frame and vice versa. For example,
if one were to have in the Jordan frame the evolution with a bounce, such that the scalar field $|\sigma| < \sqrt{12U_0}$, i.e., there is no change of the sign of the effective Newton constant,  this would mean that the behaviour of the scalar field
$\phi$ in the model with the minimally coupled scalar field and hyperbolic potential is nonsingular, and hence, the behavior of the geometric characteristics of the Universe would be nonsingular. However, as was shown  in \cite{Boisseau:2015hqa} such a behaviour does not exist, and the effective Newton constant in the Jordan frame necessarily changes sign before or after the bounce (or does it twice). This fact can be easily explained
by using the comparison with the dynamics of the corresponding model in the Einstein frame as was done in paper~\cite{Pol1}.
Indeed, in a flat Friedmann Universe filled with a minimally coupled scalar field, the time derivative of the Hubble parameter is always negative. Taking into account the explicit form of the potential (\ref{Pol-Star4}), one can conclude that this last fact is not compatible with the absence of the Big Bang or the Big Crunch or both the singularities.

\section{Conclusion: comparison between  the different approaches to the possibility of crossing the singularity}

The problem of the existence of the cosmological singularity at the beginning of  cosmological evolution has attracted the  attention of  people  studying general relativity, for a long time \cite{Khal-Lif}.   In papers \cite{Pen-Hawk}
the impossibility was shown of the indefinite continuation of geodesics under certain conditions.
This was interpreted as pointing to the existence of a singularity in the general
solution of the Einstein equations.
The analytical behaviour of the general solutions to the Einstein equations in the
neighbourhood of a singularity was investigated in papers \cite{BKL}.
These papers revealed the enigmatic  phenomenon of an oscillatory approach to the singularity
which has also become known as the {\it Mixmaster Universe} \cite{Misner}.
The model of a closed homogeneous, but anisotropic, Universe with three degrees of freedom
(Bianchi IX cosmological model) was used to demonstrate that the Universe approaches the singularity
in such a way that its contraction along two axes is accompanied by an expansion with respect to
the third axis, and the axes change their roles according to a rather complicated law which reveals
a chaotic behaviour~\cite{BKL,chaos}.

Another type of cosmological singularity, arising at finite value of the cosmological scale factor, was considered in~\cite{Barrow}.
Recently, the so called ``soft'' singularities arising for small values of the scale factor were extensively studied \cite{soft} and the
situations for which such singularities can be crossed were found \cite{cross,cross-we}. At the same time the idea that the Big Bang -- Big Crunch singularity can be crossed appeared very counterintuitive. Nonetheless,  as we frequently mentioned in the present paper, the procedure for the crossing of the Big Bang -- Big  Crunch singularity, based on the use of  Weyl symmetry,  was elaborated \cite{Bars1}--\cite{Bars7}.  Using a Weyl-invariant theory, where two conformally coupled
with gravity scalar fields were presented, the authors have obtained the geodesic completeness of the corresponding space time. The consequence  of this geodesic completeness is the crossing of the Big Bang singularity and the emergence of  antigravity regions on using the Einstein frame.
In particular, in papers \cite{Bars3,Bars5,Bars7} the
crossing of  anisotropic singularities was considered. The authors  used the
general expressions for   Bianchi-I, Bianchi-VIII and Bianchi-IX Universes. In the
vicinity of the singularity all these Universes have Kasner--like behavior. It allows one to
write down explicitly the solutions for a Universe crossing such a singularity on using some finite Weyl--invariant quantities. The explicit set of geodesics for the crossing singularity massive and massless particles  was constructed as well.

The use of Weyl symmetry to describe the passage through the Big Crunch -- Big Bang singularity accompanied by a change of sign for the effective Newton's constant has led to some discussion. In \cite{polem} it was noticed that for such a passage through the singularity some curvature invariants become infinite. In paper \cite{Bars7} a counter-argument was put forward. If one has enough conditions in order to match the nonsingular quantities before and after crossing the  singularities, then the singularities can be traversed. In the present paper, inspired by the idea of the crossing of singularities, developed in the papers cited above, and by the correspondence between the simple conformally coupled scalar field model with bounce
\cite{Boisseau:2015hqa} and the models with hyperbolic potentials \cite{Pol1,KPTVV2015}, we  propose a slightly different version of the description of the crossing of singularities. It is based on the transitions between the Jordan and Einstein frames, and here we have applied it to simpler models than those discussed previously.
As we have already emphasized in the Introduction, we only considered  an isotropic cosmological singularity present in a flat Friedmann Universe.

This research direction  seems promising, and some new papers concerning this topic have
appeared. In paper \cite{Bars8} the crossing of the Schwarzschild singularity was considered
by using a technique quite similar to that used in the analysis of the cosmological singularities. In paper~\cite{Bars9} some physical problems connected with the existence of an antigravity regime are analyzed, and the possibility of the indirect observation of such a phenomenon are discussed. Indeed, it was emphasized that on using  Weyl invariance one can get a geodesically complete theory. The price of this geodesic completeness  is the appearance of  antigravity regions of spacetime connected with gravity regions through gravitational singularities such as those that occur in black holes and cosmological bang/crunch. Antigravity regions lead to several questions regarding the physical interpretation. In \cite{Bars9} it was shown that unitarity is not violated, but there may be an instability associated with negative kinetic energies in the antigravity regions. However, this  problem can be resolved with an interpretation of the theory from the perspective of observers
living strictly in the gravity region. Such observers cannot experience the negative kinetic energy in antigravity directly but can only detect in and out signals that interact with the antigravity region.
At this point,  it is necessary to stress that the appeal to quantum cosmology in paper \cite{Bars9} was aimed not to cure the drawbacks of the classical theory but to show that even in quantum theory the problem associated with the presence of a scalar field with the wrong sign for the kinetic term, which is equivalent to the emergence of the antigravity regions, is not dangerous. Indeed, the Wheeler-DeWitt equation~\cite{WDW} for the minisuperspace model, including two scalar fields, was considered, and it was shown that the corresponding wave function of the Universe can be represented as a wave function of two harmonic oscillators
with different signs for the spectra. The model of the relativistic harmonic oscillator, which is equivalent to two oscillators with different signs for the spectra, was considered in \cite{Bars-osc}, where it was shown that the normalized wave function for such a system does exist. The fact that the spectrum of such an oscillator (or a Universe with two scalar fields and Weyl symmetry) is not bounded from below is also not dangerous if one considers   observers living in the gravity region.
Let us add another comment. It was also noticed in \cite{Bars9} that the Weyl-invariant string theory, revealing geodesic completeness, can be constructed
in a quite analogous way. This treatment of the string theory is not stimulated by the fact that in the standard gravity theory one cannot resolve some problems
and, hence, should consider the string theory. Vice versa, due to the similar structure of these two theories, the results, obtained for the Weyl generalization of standard gravity, can be easily extended to the case of the string theory.

The quantum cosmological aspects of crossing singularities were considered in paper~\cite{Turok}. The relations between different types of singularities arising in the Einstein and Jordan frames were studied in the  paper~\cite{Odin}.

A somewhat different approach to the problem of a cosmological singularity was developed in a series of papers~\cite{Wetterich1}--\cite{Wetterich5}. There the author considers the so called variable gravity together  with the transitions between different frames and the reparametrization of the scalar field. In particular, he introduces a so-called freeze frame where the Universe is very cold and slowly evolving. In the freeze picture the masses of elementary particles and the gravitational constant decrease with cosmic time, while  Newtonian attraction remains unchanged. The cosmological solution can be extrapolated to the infinite past in physical time -- the Universe has no beginning. In the equivalent, but singular Einstein frame cosmic history finds the familiar big bang description.
Generally, the papers \cite{Wetterich1}--\cite{Wetterich5} give the impression that there is no physical singularity in Nature,  but only a singularity in the choice of field coordinates. Let us note that in this scheme there is only one scalar field (just as in our present paper), but the Hilbert--Einstein term in the action is absent and the coupling coefficient between the scalar field squared and the scalar curvature is such  as to provide the positive sign of the effective Newtonian constant.
The moment of time which corresponds to the Big Bang in the Einstein frame (in other words ``in the Big Bang picture''), in the freeze frame also corresponds to the beginning of the cosmological evolution, but the geometry of the spacetime is not singular: it is the field parametrisation, which is singular. According to the definition of the author, the cosmology is
non-singular, if there is a frame, where the geometry is non-singular. Thus, there is an analogy between the horizon which arises due to a certain choice of the spacetime coordinates: the singularity arises because of some choice of the field parametrisation. Further, the evolution occupies an infinite physical time. It is stressed that the physical time does not necessarily coincide with the proper (cosmic) time. Instead, the author
of \cite{Wetterich1}--\cite{Wetterich5} defines the physical time by counting the number of zeros of a component of the wave function. Let us note that the idea that the cosmic time is not always the best parameter for the description of  physical reality also arises in the context of the oscillatory approach to the cosmological singularity \cite{BKL}. Indeed, in this situation it is better to use the logarithmic time, which is more adapt for the description of an infinite amount of Kasner epochs and eras (see, e.g., \cite{Kam-UFN} and the references therein).
The opportunity of the crossing of singularities in the Big Bang picture is not excluded in the approach \cite{Wetterich1}--\cite{Wetterich5}, but it does not arise for the concrete form of the  models considered.

On summing up, we can say that our approach is less radical. We only state that by using the field reparametrization one can describe in a unique and reasonable way a possible singularity crossing. Here, it makes sense to dwell on the question of the physical equivalence between different frame choices. Generally, one can say (see e.g. \cite{Sasaki}) that even if some classical phenomena look different in Einstein and Jordan frames, the relations between different observables should be the same. Moreover, one can state that the choice between the Einstein and the Jordan frame is somewhat similar to the choice of the frame of reference in the Newtonian mechanics. A coherent description of the laws of physics can be given in every frame of reference. Nonetheless, a natural choice
of such a frame of reference exists and is connected with each concrete physical problem.  For example, a person on merry-go round naturally describes physical phenomena around him by taking into account inertial forces such as the centrifugal and Coriolis.   For this person, the non-inertial rotating frame is the natural physical frame to be used. Analogously, in cosmology for a comoving observer the natural frame is that associated with his cosmic time (when this
is an adequate candidate for the role of the physical time, which is not always a case, as was explained in the preceding paragraph). Thus, since the transition between the Einstein and the Jordan frames implies the change of the cosmic time, the observed cosmological evolutions in these frames are different.

Let us  once again look at the analogy between the appearance of  horizons as the result of a certain choice of the spacetime coordinates and the appearance of the singularity as the result of a choice of the fields parametrization. Even if the appearance of the horizon in the Schwarzschild metric is a result of a standard choice of the coordinates, its consequences for a distant observer are quite real: he cannot get the signals, coming from the region which stays behind the horizon. The introduction of  Kruskal coordinates \cite{Kruskal}, which makes the spacetime manifold geodesic complete does not eliminate this effect of the horizon. However, the Kruskal coordinates tell us what happens behind the horizon, in the region which is unaccessible for us. Thus, from our point of view for certain observers, the singularity indeed exist and the transition to the frame where it does not exist, allows us to describe in a unique way the passage through the singularity. At this
point, we can say that in the passage through a singularity of the Big Bang -- Big Crunch type all the extended objects are destroyed, in spite of the fact that in another frame, the value of the scale factor is finite.

It is important to note that the possibility of a change of sign of the effective gravitational  constant in the model with the scalar field conformally coupled with the scalar curvature was studied in paper \cite{Starobinsky1981}, where the earlier suggestion, made in paper \cite{Linde1979} was analyzed in some detail. In paper \cite{Starobinsky1981} it was pointed out  that in a homogeneous and isotropic Universe, one can indeed  cross the point where the effective gravitational constant changes sign.
However, the presence of anisotropies or inhomogeneities changes the situation drastically, because when the value of the effective gravitational constant tends to zero, these anisotropies and inhomogeneities grow indefinitely. It seems to us that this phenomenon can be easily understood in the framework developed in the present paper. Let us consider, for example, the Bianchi-I Universe in the presence of the minimally coupled scalar field. In the vicinity of the singularity the metric contains isotropic and anisotropic parts. On making the conformal transformation to the Jordan frame, combined with the reparametrization of the scalar field, we  obtain the model with a conformally coupled scalar field. As was explained in detail above, such a transformation leads us to the situation wherein the function $U_c$ tends to zero in the Jordan frame, while the homogeneous part of the metric,
described by the scale factor evolves quite  regularly in this frame. However, this conformal transformation acts only on the  isotropic part of the metric, leaving intact the anisotropic factors, which are the same in the Jordan and in the Einstein frames, and
hence, are singular also in the Jordan frame. Thus, the mechanism  which we have used to describe the crossing  of  the singularity
in one frame,  by using the conformal transformation to another frame in the  Friedmann Universes, should be modified if we wish to consider the anisotropic or inhomogeneous models.
Let us mention once again that in papers \cite{Bars3,Bars5,Bars7} the problem of passing through an anisotropic singularity was solved in the framework of the Weyl-invariant gravity theory with two scalar fields.  The question of possibility of the description of the crossing of  such a singularity in our approach requires further investigation~\cite{future}.

In paper \cite{Brand} the inhomogeneous cosmologies in the models \cite{Bars1}--\cite{Bars7} were studied, and it was argued that a consistent treatment of  cosmological perturbations in the presence of the transition between gravity and antigravity regimes is possible. The relations between the cosmological perturbations
in the Einstein and Jordan frames were studied in paper \cite{Qiu}. The inhomogeneous cosmological models in the presence of a non-minimally coupled scalar field were studied in \cite{Caputa}. There it was stressed that while the homogeneous and isotropic cosmology appears surprisingly sensible when the effective Planck mass vanishes, the small inhomogeneities provoke  catastrophical consequences. In paper \cite{Brand} the authors stated that there is an essential difference between the model with two scalar fields \cite{Bars1}--\cite{Bars7} and a model with one non-minimally coupled scalar field, because in the former one can avoid the appearance of the singularities which arise in the latter and which were studied in \cite{Starobinsky1981} and \cite{Caputa} and in some other works \cite{other-anis-inh}.   However, we think that effectively one of the scalar degrees of freedom is used
to fix a  gauge choice and in this sense the description in terms of two fields, used in  \cite{Bars1}--\cite{Bars7} and \cite{Brand} is very close to the description in terms of  one scalar field. Indeed, in the description of the inhomogeneities on the Friedman--Robertson--Walker background the authors reduce the scalar perturbations to one function. Thus, we hope to develop the mechanism of the continuation of the scalar perturbations through the singularity for our models too~\cite{future}.

Let us emphasize that
 the continuation through the singularity is possible  when we can make a unequivocal matching between the geometries on the different sides of the singularity. In the case of Bianchi Universes the parts of the metric, describing anisotropies can be characterized
 by some finite constants, in spite of the fact that the corresponding curvature invariants
 diverge when the spatial volume of the Universe tends to zero. The existence of such constants allows the description of the continuation of the geometry through the singularity.
 The treatment of inhomogeneities is more complicated but, again, is not hopeless.
 We hope to study these topics in the forthcoming work~\cite{future}.

To conclude, we wish to say that in all three  approaches (two-scalar-field approach \cite{Bars1}--\cite{Bars7}, variable gravity approach \cite{Wetterich1}--\cite{Wetterich5} and the Jordan--Einstein frames transformation approach proposed in the present paper)  for the description of the passage through the Big Bang -- Big Crunch type singularities, Weyl symmetry was used.
In papers \cite{Bars1}--\cite{Bars7} two scalar fields conformally coupled with gravity  were introduced and, by fixing the gauge in different ways, the authors showed how it is possible to describe singularity crossing. Here, in order to reach the same goal, we have used the
transformations between the Jordan and Einstein frames and in the Jordan frame the coupling was the conformal one. We have used a conformal coupling because in this case the relations between the parametrizations of the scalar field in the Jordan and in the Einstein frame have a simple explicit form. It is not clear to us if it is a purely technical expedient or there is something deep behind this introduction of Weyl symmetry.
One can ask another question: Is it possible to find a method which describes singularity crossing in cosmological models wherein a scalar field is absent, for example in models of Universes filled with fluids?
Further, the very presence of matter different from a scalar field can change the regime of crossing the singularity; such effects were studied  for the case of sudden future singularities \cite{cross-we}. In this case there is no general recipe and one has to explore the situation case by case.
Lastly, we wish to cite the paper \cite{Awad}, where the Weyl-anomaly quantum correction to the Friedmann equations
was taken into account, thus transforming the Big Bang -- Big Crunch singularity in a soft traversable future singularity. This confirms that
the study of cosmological singularities at the level of homogeneous and isotropic Universes  is a topic of active interest.
We believe that all these questions deserve further study, even if at a first glance they may appear a bit exotic.

\section*{Acknowledgments}
We  thank C.~Wetterich for useful communications.
Research of E.P. is supported in part by Grant No.~MK-7835.2016.2  of the President of Russian Federation.
Research of E.P. and S.V. is supported in part by the RFBR Grant No.~14-01-00707. A.K. was partially supported by the RFBR Grant No.~14-02-00894.


\begin{thebibliography}{99}
\bibitem{inflation}
A.A.~Starobinsky,
Lect. Notes Phys.  {\bf 246}, 107 (1986);
A.D.~Linde,  \textit{Particle Physics and Inflationary Cosmology}, (Harwood, Chur, Switzerland, 1990).
\bibitem{cosmic}
A.~Riess \textit{et al.},
 Astron. J.  {\bf 116},  1009 (1998);
S.J.~Perlmutter \textit{et al.},
Astrophys. J.  {\bf 517},  565 (1999).
\bibitem{dark}
V. Sahni V and A.A. Starobinsky,
Int. J. Mod. Phys. D  {\bf 9},  373  (2000);
T.~Padmanabhan,
 Phys. Rep. {\bf 380},  235  (2003);
P.J.E.~Peebles and B.~Ratra,
Rev.\ Mod.\ Phys.\ {\bf 75}, 559 (2003);
V. Sahni,
Class. Quantum Grav. {\bf 19},  3435  (2002);
E.J. Copeland, M. Sami  and S. Tsujikawa,
Int. J. Mod. Phys. D {\bf 15},  1753  (2006);
V.~Sahni  and A.A.~Starobinsky,
Int. J. Mod. Phys.  D {\bf 15},  2105 (2006);
S. Tsujikawa,
Class. Quantum Grav. {\bf 30}, 214003 (2013).
\bibitem{BI}
 M.~Born and L.~Infeld,
  Proc.\ Roy.\ Soc.\ Lond.\ A {\bf 144}, 425 (1934).
\bibitem{tach}
A.~Sen,
  JHEP {\bf 0204}, 048 (2002).
\bibitem{k}
C.~Armendariz-Picon, V.F.~Mukhanov and P.J.~Steinhardt,
  Phys.\ Rev.\ D {\bf 63}, 103510 (2001).
\bibitem{Jordan}
P.~Jordan, \textit{Schwerkraft und Weltall}, Vieweg (Braunschweig) 1955.
\bibitem{Sakharov}
A.D.~Sakharov,
 Sov. Phys. Dokl. {\bf 12},   1040 (1967).
\bibitem{induced}
F.~Cooper and G.~Venturi,
Phys. Rev.  D {\bf 24},   3338 (1981);
F.~Finelli, A.~Tronconi  and G.~Venturi,
Phys. Lett.  B  {\bf 659},  466  (2008);
A.~Cerioni, F.~Finelli, A.~Tronconi  and G.~Venturi,
Phys. Lett.  B {\bf 681},  383  (2009);
A.~Cerioni, F.~Finelli, A.~Tronconi  and G.~Venturi,
Phys. Rev.  D {\bf 81},  123505 (2010);
A.~Tronconi  and G.~Venturi,
Phys. Rev.  D {\bf 84},  063517 (2011);
 A.Yu.~Kamenshchik, A.~Tronconi  and G.~Venturi,
Phys. Lett.  B {\bf 713},  358 (2012).

\bibitem{ChernikovTagirov}
  N.A.~Chernikov,  E.A.~Tagirov,
  Annales Poincare Phys.\ Theor.\ A  \textbf{9},  109 (1968);
  E.A.~Tagirov,
  Annals Phys. \textbf{76}, 561 (1973)

\bibitem{Callan:1970ze}
  C.G.~Callan, S.R.~Coleman,  R.~Jackiw,
  Annals Phys. \textbf{59} (1970) 42

\bibitem{nonmin-inf}
B.L.~Spokoiny,
Phys. Lett. B {\bf 147},  39 (1984);
T.~Futamase  and K.I.~Maeda,
Phys. Rev.  D {\bf 39},  399 (1989);
D.S.~Salopek, J.R.~Bond  and  J.M.~Bardeen,
 Phys. Rev.  D {\bf 40}, 1753 (1989);
R.~Fakir  and W.G.~Unruh,
Phys. Rev.  D {\bf 41},  1783 (1990).
\bibitem{nonmin-quant}
A.O.~Barvinsky  and A.Yu.~Kamenshchik,
Phys. Lett.  B {\bf 332},  270  (1994);
A.O.~Barvinsky, A.Yu.~Kamenshchik, C.~Kiefer  and C.F.~Steinwachs,
 Phys. Rev. D {\bf 81}, 043530 (2010);
 A.O.~Barvinsky, A.Y.~Kamenshchik and D.V.~Nesterov,
  Eur.\ Phys.\ J.\ C {\bf 75}, no. 12, 584 (2015).

\bibitem{Higgs}
F.L.~Bezrukov   and M.~Shaposhnikov,
Phys. Lett. B {\bf 659}, 703 (2008);
A.O. Barvinsky, A.Yu. Kamenshchik  and A.A. Starobinsky,
J. Cosmol. Astropart. Phys. {\bf 0811},  021 (2008);
F.L. Bezrukov, A. Magnin and M. Shaposhnikov,
Phys. Lett. B {\bf 675},  88 (2009);
A. De~Simone A, M.P. Hertzberg  and F. Wilczek,
Phys. Lett.  B {\bf 678}, 1 (2009);
A.O. Barvinsky, A.Yu. Kamenshchik, C. Kiefer, A.A. Starobinsky  and C. Steinwachs,
J. Cosmol. Astropart. Phys. {\bf 0912}, 003 (2009);
F.L.~Bezrukov, A.~Magnin, M.~Shaposhnikov  and S.~Sibiryakov,
 J. High Energy Phys. {\bf 1101},  016 (2011);
A.O.~Barvinsky, A.Yu.~Kamenshchik, C. Kiefer, A.A.~Starobinsky  and C.F.~Steinwachs,
Eur. Phys. J. C {\bf 72},  2219 (2012);
F.~Bezrukov  and D.~Gorbunov,
J. High Energy Phys. {\bf 1307},  140 (2013);
F.~Bezrukov,
 Class. Quant. Grav. {\bf 30},  214001 (2013).
\bibitem{Wagoner}
R.W.~Wagoner,
 Phys. Rev. D {\bf 1},  3209 (1970).

\bibitem{debate}
V.~Faraoni and E.~Gunzig,
  Int.\ J.\ Theor.\ Phys.\  {\bf 38}, 217 (1999);
  R.~Catena, M.~Pietroni and L.~Scarabello,
  Phys.\ Rev.\ D {\bf 76}, 084039 (2007);
  S.~Capozziello, P.~Martin-Moruno and C.~Rubano,
  Phys.\ Lett.\ B {\bf 689}, 117 (2010);
  C.F.~Steinwachs and A.Y.~Kamenshchik,
  Phys.\ Rev.\ D {\bf 84}, 024026 (2011);
  X.~Calmet and T.C.~Yang,
  Int.\ J.\ Mod.\ Phys.\ A {\bf 28}, 1350042 (2013);
    Y.N.~Obukhov and D.~Puetzfeld,
  Phys.\ Rev.\ D {\bf 90},  104041 (2014);
G.~Calcagni, C.~Kiefer and C.F.~Steinwachs,
  J. Cosmol. Astropart. Phys. {\bf 1410}, no. 10, 026 (2014);
A.Y.~Kamenshchik and C.F.~Steinwachs,
  Phys.\ Rev.\ D {\bf 91}, 084033 (2015);
L.~Jarv, P.~Kuusk, M.~Saal and O.~Vilson,
  Phys.\ Rev.\ D {\bf 91}, 024041 (2015);
G.~Domenech and M.~Sasaki,
  J. Cosmol. Astropart. Phys. {\bf 1504}, no. 04, 022 (2015);
  O.~Hrycyna and M.~Szydlowski,
  J. Cosmol. Astropart. Phys. {\bf 1511}, no. 11, 013 (2015);
I.G.~Moss,
  arXiv:1509.03554 [hep-th];
M.~Herrero-Valea,
 Phys. Rev. D \textbf{93}, 105038 (2016),
\bibitem{Sasaki}
N.~Deruelle and M.~Sasaki,
  Springer Proc.\ Phys.\  {\bf 137}, 247 (2011).
\bibitem{Fre}
 P.~Fr\'e, A.~Sagnotti  and A.S.~Sorin,
Nucl. Phys. B {\bf 877},  1028 (2013).
\bibitem{KPTVV2013}
A.Yu.~Kamenshchik, E.O.~Pozdeeva, A.~Tronconi, G.~Venturi and S.Yu.~Vernov,
  Class.\ Quant.\ Grav.\  {\bf 31},  105003 (2014).
\bibitem{we-ind-ex}
A.Yu.~Kamenshchik,  A.~Tronconi  and G.~Venturi,
Phys. Lett. B {\bf 702},  191 (2011).
\bibitem{exp-part}
F.~Lucchin  and S.~Matarrese,
Phys. Rev. D {\bf 32},  1316 (1985);
J.J.~Halliwell,
Phys. Lett. B {\bf 185},  341 (1987);
 J.D.~Barrow,
 Phys. Lett. B {\bf 187}, 12 (1987);
A.B.~Burd   and J.D.~Barrow,
Nucl. Phys. B {\bf 308}  929 (1988);
G.F.R.~Ellis  and M.S.~Madsen,
Class. Quantum Grav. {\bf 8},  667 (1991);
V.~Gorini, A.Yu.~Kamenshchik, U.~Moschella  and V.~Pasquier,
Phys. Rev. D {\bf 69}, 123512 (2004).
\bibitem{Gasp}
M.~Gasperini and G.~Veneziano,
  Phys.\ Rept.\  {\bf 373}, 1 (2003).
\bibitem{Pol1}
B.~Boisseau, H.~Giacomini and D.~Polarski,
  J. Cosmol. Astropart. Phys. {\bf 1510},  033 (2015).
\bibitem{Boisseau:2015hqa}
  B.~Boisseau, H.~Giacomini, D.~Polarski and A.A.~Starobinsky,
  J. Cosmol. Astropart. Phys. {\bf 1507}, 002 (2015).


\bibitem{Bars1}
I.~Bars and S.H.~Chen,
  Phys.\ Rev.\ D {\bf 83}, 043522 (2011).
  \bibitem{Bars2}
  I.~Bars, S.H.~Chen and N.~Turok,
  Phys.\ Rev.\ D {\bf 84}, 083513 (2011).
\bibitem{Bars3}
I.~Bars, S.H.~Chen, P.J.~Steinhardt and N.~Turok,
  Phys.\ Lett.\ B {\bf 715}, 278 (2012).
  \bibitem{Bars4}
  I.~Bars, S.H.~Chen, P.J.~Steinhardt and N.~Turok,
  Phys.\ Rev.\ D {\bf 86}, 083542 (2012).
  \bibitem{Bars5}
  I.~Bars,
  arXiv:1209.1068 [hep-th].
\bibitem{Bars6}
I.~Bars, P.~Steinhardt and N.~Turok,
  Phys.\ Rev.\ D {\bf 89},  043515 (2014).
  \bibitem{Bars7}
  I.~Bars, P.~Steinhardt and N.~Turok,
  Phys.\ Rev.\ D {\bf 89},  061302 (2014).



\bibitem{soft}
Y.~Shtanov and V.~Sahni,
  Class.\ Quant.\ Grav.\  {\bf 19}, L101 (2002);
  J.D.~Barrow,
  Class.\ Quant.\ Grav.\  {\bf 21}, L79 (2004);
  J.D.~Barrow,
  Class.\ Quant.\ Grav.\  {\bf 21}, 5619 (2004);
  S.~Nojiri, S.D.~Odintsov and S.~Tsujikawa,
  Phys.\ Rev.\ D {\bf 71}, 063004 (2005);
  M.P.~Dabrowski and T.~Denkiewicz,
  Phys.\ Rev.\ D {\bf 79}, 063521 (2009);

\bibitem{cross}
L.~Fernandez-Jambrina and R.~Lazkoz,
  Phys.\ Rev.\ D {\bf 70}, 121503 (2004)
\bibitem{cross-we}
Z.~Keresztes, L.A.~Gergely, A.Y.~Kamenshchik, V.~Gorini and D.~Polarski,
  Phys.\ Rev.\ D {\bf 82}, 123534 (2010);
 Z.~Keresztes, L.~A.~Gergely and A.~Y.~Kamenshchik,
  Phys.\ Rev.\ D {\bf 86}, 063522 (2012);

   Z.~Keresztes, L.A.~Gergely, A.Y.~Kamenshchik, V.~Gorini and D.~Polarski,
  Phys.\ Rev.\ D {\bf 88}, 023535 (2013);
  A.Y.~Kamenshchik,
  Class.\ Quant.\ Grav.\  {\bf 30}, 173001 (2013).
\bibitem{polem}
J.J.M.~Carrasco, W.~Chemissany and R.~Kallosh,
  JHEP {\bf 1401}, 130 (2014);
  R.~Kallosh and A.~Linde,
  J. Cosmol. Astropart. Phys. {\bf 1401}, 020 (2014).

\bibitem{Roest}
  M.~Ozkan and D.~Roest,
  arXiv:1507.03603 [hep-th].


\bibitem{KPTVV2015}
  A.Yu.~Kamenshchik, E.O.~Pozdeeva, A.~Tronconi, G.~Venturi and S.Yu.~Vernov,
  Class. Quantum Grav. {\bf 33}, 015004 (2016).

\bibitem{Bars8}
  I.J.~Araya, I.~Bars and A.~James,
  arXiv:1510.03396 [hep-th].

\bibitem{Bars9}
I.~Bars and A.~James,
Phys. Rev. D {\bf 93},  044029 (2016)



\bibitem{Wetterich1}
C.~Wetterich,
  Phys.\ Rev.\ D {\bf 89},  024005 (2014).
\bibitem{Wetterich2}
C.~Wetterich,
  Phys.\ Rev.\ D {\bf 90},  043520 (2014).
  \bibitem{Wetterich3}
C.~Wetterich,
  Phys.\ Lett.\ B {\bf 736}, 506 (2014).
  \bibitem{Wetterich4}
C.~Wetterich,
  Nucl.\ Phys.\ B {\bf 897}, 111 (2015).
  \bibitem{Wetterich5}
C.~Wetterich,
    J. Cosmol. Astropart. Phys. \textbf{1605}, 041 (2016)

\bibitem{KKhT}
A.Y.~Kamenshchik, I.M.~Khalatnikov,  A.V.~Toporensky,
  Int.\ J.\ Mod.\ Phys.\ D  \textbf{6}, 673 (1997).

\bibitem{ABGV}
I.Ya.~Aref'eva, N.V.~Bulatov, R.V.~Gorbachev, S.Yu.~Vernov,
Class. Quant. Grav. \textbf{31}, 065007 (2014).

\bibitem{Pozdeeva2014}
E.O.~Pozdeeva and S.Yu.~Vernov,
  AIP Conf. Proc. \textbf{1606},  48 (2014).

\bibitem{AJV2} I.Ya.~Aref'eva, L.V.~Joukovskaya, and S.Yu.~Vernov,
 J. Phys. A: Math. Theor. \textbf{41}, 304003 (2008).

\bibitem{KTVV2013}
A.Yu.~Kamenshchik, A.~Tronconi, G.~Venturi, and S.Yu.~Vernov,
Phys. Rev. D \textbf{87},  063503 (2013).

\bibitem{Kruskal}
M.D.~Kruskal,
  Phys.\ Rev.\  {\bf 119}, 1743 (1960).

\bibitem{Khal-Lif}
E.~M.~Lifshitz and I.~M.~Khalatnikov,
  Adv.\ Phys.\  {\bf 12}, 185 (1963).

\bibitem{Pen-Hawk}
R.~Penrose, {\it Structure of Space-Time} (New York, Amsterdam:W.A. Benjamin, 1968);
S.W. Hawking and G.F.R.  Ellis, {\it The Large Scale Structure of Space-Time}
(Cambridge, New York: Cambridge Unive. Press, 1973);
S.W. Hawking and R.  Penrose,  Proc. R. Soc. London Ser. A {\bf 314}, 529 (1970).

\bibitem{BKL}
V.~A.~Belinsky, I.~M.~Khalatnikov and E.~M.~Lifshitz,
  Adv.\ Phys.\  {\bf 19}, 525 (1970); {\bf 31} 639 (1982).
\bibitem{Misner}
C.W. Misner,  Phys. Rev. Lett {\bf 22}, 1071 (1969).
\bibitem{chaos}
E.M.~Lifshitz, I.M.~Lifshitz and I.M.~Khalatnikov,  Sov. Phys. JETP
{\bf 32}, 173 (1971);
I.M.~Khalatnikov, E.M.~Lifshitz, K.M.~Khanin, L.N.~Shchur and Ya.G.~Sinai,  J. Stat. Phys. {\bf 38}, 97 (1985).
\bibitem{Barrow}
J.D.~Barrow, G.J.~Galloway and F.J.~Tipler, Mon. Not. R. Astron.
Soc. { \bf 223}, 835 (1986).
\bibitem{WDW}
B.S.~DeWitt,
  Phys.\ Rev.\  {\bf 160}, 1113 (1967).
\bibitem{Bars-osc}
I.~Bars,
  Phys.\ Rev.\ D {\bf 79}, 045009 (2009).
\bibitem{Turok}
S.~Gielen and N.~Turok,
   	Phys. Rev. Lett. \textbf{117}, 021301 (2016)
\bibitem{Odin}
S.D.~Odintsov and V.K.~Oikonomou,
  arXiv:1512.04787 [gr-qc].
\bibitem{Kam-UFN}
A.~Y.~Kamenshchik,
  Phys.\ Usp.\  {\bf 53}, 301 (2010).
\bibitem{Starobinsky1981}
A.A.~Starobinsky,
Sov. Astron. Lett. \textbf{7}, 36 (1981).

\bibitem{Linde1979}
A.D.~Linde, JETP Lett. {\bf 30}, 447 (1980).
\bibitem{future}
A.Yu.~Kamenshchik, E.O.~Pozdeeva, A.~Tronconi, G.~Venturi and S.Yu.~Vernov,
 work in progress.
\bibitem{Brand}
M.~Oltean and R.~Brandenberger, Phys. Rev. D {\bf 90}, 083505 (2014).
\bibitem{Qiu}
T.~Qiu,
  J. Cosmol. Astropart. Phys.  {\bf 1206}, 041 (2012).
\bibitem{Caputa}
P.~Caputa, S.S.~Haque, J.~Olson and B.~Underwood,
  Class.\ Quant.\ Grav.\  {\bf 30}, 195013 (2013).
\bibitem{other-anis-inh}
T. Futamase and K.i. Maeda, Phys. Rev. D {\bf 39}, 399 (1989); T. Morishima and T. Futamase, Phys. Lett. B {\bf 447}, 46 (1999); K. Bronnikov and Y. Kireev, Phys. Lett. A {\bf 67}, 95 (1978); T. Futamase, T. Rothman and R. Matzner, Phys. Rev. D {\bf 39}, 405 (1989).

\bibitem{Awad}
A.~Awad,
  Phys.\ Rev.\ D {\bf 93},  084006 (2016).

\end{thebibliography}
\end{document}